\documentclass[journal]{IEEEtran}

\ifCLASSINFOpdf
\else
\fi
\usepackage{amsmath}    
\usepackage{amssymb}    
\usepackage{bm}         
\usepackage{booktabs}
\usepackage{booktabs}
\newcommand{\doublemidrule}{%
  \specialrule{\lightrulewidth}{0pt}{0pt}%
  \specialrule{\lightrulewidth}{-0.35ex}{0.6ex}%
}

\usepackage{graphicx}
\usepackage{cite}
\usepackage[colorlinks=true, allcolors=black]{hyperref}

\usepackage[capitalise,noabbrev,nameinlink]{cleveref}
\crefname{figure}{Fig.}{Figs.}
\Crefname{figure}{Figure}{Figures}
\crefname{equation}{Eq.}{Eqs.}
\Crefname{equation}{Equation}{Equations}
\crefname{table}{Table}{Tables}
\Crefname{table}{Table}{Tables}
\crefname{section}{Sec.}{Secs.}
\Crefname{section}{Section}{Sections}

\IEEEaftertitletext{\vspace{-1.5\baselineskip}} 

\usepackage{graphicx}
\usepackage{placeins}

\begin{document}
\title{Multivariable Current Controller for Enhancing Dynamic Response and Grid Synchronization Stability of IBRs}
%
\author{Hassan~Yazdani,~Ali~Maleki,~Saeed~Lotfifard, and~Ali~Saberi
\thanks{H. Yazdani, A. Maleki, S. Lotfifard, and A. Saberi are with the School of Electrical Engineering and Computer Science, Washington State University, Pullman, WA 99164 USA (e-mail: hassan.yazdani@wsu.edu; ali.maleki@wsu.edu; s.lotfifard@wsu.edu; saberi@wsu.edu).}%
}
\maketitle
\begin{abstract}
This paper develops a multivariable current control strategy for inverter-based resources (IBRs) based on optimal control theory to enhance their dynamic performance and grid synchronization stability. The structure of the implemented multiple-input, multiple-output (MIMO) controller closely resembles that of the commonly used conventional single-input, single-output (SISO) PI controllers for IBRs. As a result, it requires only minor adjustments to conventional vector current control schemes, thereby facilitating its straightforward adoption. Time-domain simulations and analytical analysis demonstrate the superior performance of the developed method under various conditions and use case scenarios, such as weak power systems and uncertain parameters.
\end{abstract}


\section*{Nomenclature}
\subsection{Abbreviations}
\addcontentsline{toc}{section}{Nomenclature}
\begin{IEEEdescription}[\IEEEusemathlabelsep
  \IEEEsetlabelwidth{$\mathbf{u}^{*}=\lbrack v_{id}^{*},\,v_{iq}^{*}\rbrack$}]
\item[IBR] Inverter-based resource.
\item[LQR] Linear quadratic regulator.
\item[MIMO] Multi-input multi-output.
\item[PLL] Phase-locked loop.
\item[SISO] Single-input single-output.
\item[SCR] Short circuit ratio.
\item[VSC] Voltage source converter.
\end{IEEEdescription}
\vspace{-0.8\baselineskip}
\subsection{Parameters}
\begin{IEEEdescription}[\IEEEusemathlabelsep
  \IEEEsetlabelwidth{$\mathbf{u}^{*}=\lbrack v_{id}^{*},\,v_{iq}^{*}\rbrack$}]
\item[$v_{id}, v_{iq}$] $dq$-axis input voltage of the VSC.
\item[$v_{od}, v_{oq}$] $dq$-axis output voltage of the VSC.
\item[$v_{gd}, v_{gq}$] $dq$-axis Thevenin voltage of the grid.
\item[$R_f, L_f, C_f$] Filter parameters of the VSC.
\item[$R_g, L_g, Z_g$] Thevenin impedance of the grid.
\item[$i_{id}, i_{iq}$] $dq$-axis input current of the VSC.
\item[$i_{od}, i_{oq}$] $dq$-axis current of the grid.
\item[$\mathbf{A}, \mathbf{B}, \mathbf{C}$] Original system’s state space.
\item[$\overline{\mathbf{A}}, \overline{\mathbf{B}}, \overline{\mathbf{C}}$] Setpoints of $dq$-axis input current.
\item[$x^{*}=\lbrack i_{id}^{*},\,i_{iq}^{*}\rbrack$] Setpoints of $dq$-axis input current.
\item[$u^{*}=\lbrack v_{id}^{*},\,v_{iq}^{*}\rbrack$] Constant inputs corresponding to $x^{*}$.
\end{IEEEdescription}
\vspace{-0.8\baselineskip}
\subsection{Controller parameters}
\begin{IEEEdescription}[\IEEEusemathlabelsep
  \IEEEsetlabelwidth{$\mathbf{u}^{*}=\lbrack v_{id}^{*},\,v_{iq}^{*}\rbrack$}]
\item[$z(t)$] Integral state of the PI controller.
\item[$k_{p}, k_{i}$] PI gains of the SISO current controller.
\item[$k_{p}^{PLL},\, k_{i}^{PLL}$] PLL controller gains.
\item[$\mathbf{K},\, \mathbf{K}_{\mathrm{P}},\, \mathbf{K}_{\mathrm{I}}$] Optimal PI controller gains.
\item[$\overline{\mathbf{Q}},\, \overline{\mathbf{R}}$] State and input penalty of the LQR.
\item[$J,\, \overline{\mathbf{P}}$] LQR regulator and the Riccati solution.

\end{IEEEdescription}

\IEEEpeerreviewmaketitle

\section{Introduction}
\IEEEPARstart{P}{ower} grids are adopting a higher share of inverter-based resources (IBR), such as wind and solar power generation and battery energy storage systems. As a result, the dynamics of IBRs are becoming influential on the overall power grid dynamic responses, making it crucial to enhance the IBR controller's dynamics \cite{Ghiasi2023,Saeedian2022,Lliuyacc2017,AvilaMartinez2024,DiazSanahuja2022, zhang2026novel}.

One of the most common control strategies for the voltage source converter (VSC) is vector control in the synchronous reference frame (SRF) \cite{Lu2018}. The model of the system in the $dq$ frame constitutes a MIMO system \cite{deSouza2023multivariable}. The $d$ and $q$ axes are decoupled in conventional controllers using ideal proportional decoupling terms. Then, SISO PI controllers are used to control each axis. This control approach is effective and easy to implement, making it a popular controller for VSCs. However, it has limitations regarding dynamic performance and grid synchronization stability. 

Due to various neglected dynamics in designing the decoupling terms in conventional controllers, such as delays in different components of the inverter and the phase-locked loop (PLL) dynamics, the two axes are not fully decoupled during transients. During disturbances, especially in weak grid conditions, the response of one axis is affected by the other. This deteriorates the dynamic response of the VSC when facing disturbances in the power grid or changes in the controller's set points.

Several attempts have been made to address the above challenges. Complex transfer functions were initially used for designing induction motors' current controllers \cite{deAguiar2000, delBlanco1999}. The VSC's transfer function symmetry is leveraged in the complex transfer function to convert the original MIMO transfer functions into SISO transfer functions. In this way, a conventional SISO controller can be used for the VSC, providing better responses than directly using conventional ideal decoupling terms \cite{Yazdanian2014, Bahrani2011}. However, utilizing a complex transfer function restricts the degree of freedom in controller design, as the controllers for different axes of the original MIMO system cannot be designed independently. Also, direct cancelation of zeros of controllers and poles of the VSCs is prone to numerical inaccuracies and model uncertainties.  

In \cite{Silwal2022}, a full-state feedback MIMO controller is developed where the performance is enhanced by incorporating the PLL dynamics in the controller design process. Accessing the full state of the system may require additional state observers, which complicates the control design. In \cite{Antoine2021}, PLL dynamics are also added to the state space model of the system used for designing the controller. Such explicit modeling of the PLL dynamic in the state space requires information about the power grid Thevenin equivalent parameters, which is a challenging task to obtain. The Thevenin equivalent may also change over time due to changes in the power grid's operating point and/or topology.

Nonlinear controllers such as feedback linearization and sliding mode controllers \cite{Shao2022} and Lyapunov-based controller design and stability analysis \cite{Fan2022, Pal2022} have also been proposed to enhance the responses of VSC current controllers. However, such methods do not preserve the structure of the conventional VSC, which increases the complexity of the control design, tuning, and studying system impacts.

Different control parameter tuning strategies have also been proposed, developing trade-offs among controller objectives. For instance, \cite{Li2022TPWRS} recommends reducing the bandwidth of controller parameters to enhance the grid synchronization stability. Although this method reduces the interactions between the control loops, it may lead to a sluggish controller that cannot track the fast-changing condition of the system \cite{Zhou2014, Huang2018}. 

This paper implements a MIMO current control approach based on optimal PI control theory to enhance the response of VSCs with the following features:
\begin{itemize}
    \item The structure of the controller resembles that of the commonly used conventional PI-based VSC controllers. It does not add any complexity to the controller structure as it preserves the structure of conventional controllers while offering superior performance. 
    \item The controller parameter tuning is systematic and utilizes optimal control design using linear quadratic regulator (LQR) which is a well-established control synthesis method. As such it has classical robustness property (i.e., gain and phase margin) associated with LQR design methodology.
    \item The controller enhances the dynamics response of VSCs by providing a faster response, less overshoot, and faster settling time. 
    \item The controller improves the grid synchronization stability of VSCs. This is demonstrated through time-domain simulations and analytical analysis, where it does not require rigorous tuning as the grid strength changes \cite{Li2022TPWRS}. 
\end{itemize}

The rest of the paper is organized as follows: Section II briefly explains the structure of conventional SISO current controllers. Section III explains the developed current controller for VSCs. Section IV presents time-domain results and analytical analysis that demonstrate enhanced dynamic performance and grid synchronization stability of the controller. Finally, section V concludes the paper.
\begin{figure}[!t]
  \centering
  \includegraphics[width=\linewidth]{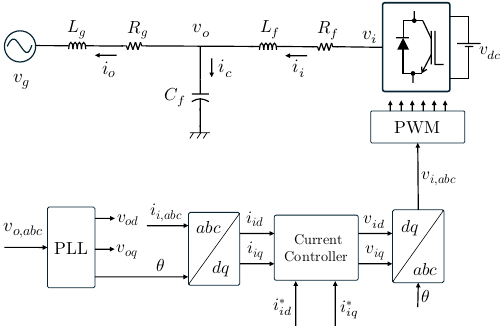}
  \caption{Schematics of a typical controller of VSC.}
  \label{fig:GFL}
\end{figure}
\section{Conventional SISO Current Controller}
\cref{fig:GFL} shows a typical grid connected IBR. By applying Kirchhoff's Voltage Law (KVL) to the filter, the following can be derived: 

\begin{equation}
\label{eq:diid/dt}
\frac{di_{id}}{dt} = \frac{1}{L_f}(v_{id} - R_f i_{id} + \omega L_f i_{iq} - v_{od})
\end{equation}
\begin{equation}
\label{eq:diiq/dt}
\frac{di_{iq}}{dt} = \frac{1}{L_f}(v_{iq} - R_f i_{iq} - \omega L_f i_{id} - v_{oq})
\end{equation}
\begin{figure}[!t]
  \centering
  \includegraphics[width=\linewidth]{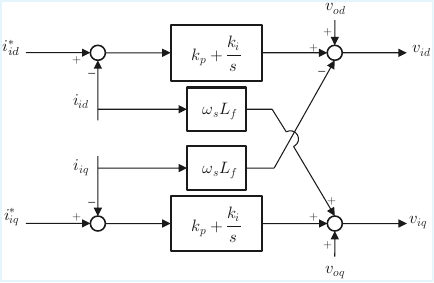}
  \caption{Conventional SISO-PI based current controller of VSCs.}
  \label{fig:SISO_PI}
\end{figure}
In \eqref{eq:diid/dt} and \eqref{eq:diiq/dt}, $v_{id}$ and $v_{id}$ are input variables and $i_{id}$ and $i_{iq}$ are output variables. The VSC controller determines the values of $v_{id}$ and $v_{iq}$ such that $i_{id}$ and $i_{iq}$ become equal to the desired values determined by the set points $i_{id}^{*}$ and $i_{iq}^{*}$. 

According to \eqref{eq:diid/dt}, the state space in $d$-axis is coupled with that of $q$-axis due to the $\omega L_{f} i_{iq}$ term. Similarly, according to \eqref{eq:diiq/dt}, the state space in $q$-axis is coupled with that of $d$-axis due to the $-\omega L_f i_{id}$ term. In conventional vector controllers of VSC, these two terms are measured and provided to the controllers of each axis to cancel out their impacts. Once two axes are decoupled, conventional SISO PI can be used to control each axis. Note that $v_{od}$ and $v_{oq}$ are also measured and are provided as feedback to the controller to cancel out their impacts. \cref{fig:SISO_PI} shows the overall schematic of the conventional current controller.

Using the above-mentioned ideal decoupling terms (i.e., $-\omega_{s} L_f i_{iq}$ and $\omega_{s} L_fi_{iq}$) has an acceptable steady-state response, but it deteriorates the transient response of the controller. This is because due to delays in the measurement filters, PWM, and the dynamic response of the PLL or any other grid synchronization mechanisms the ideal cancelation of the coupling terms is impossible. 

\section{MIMO Current Controller for VSC}
\begin{figure}[!t]
  \centering
  \includegraphics[width=\linewidth]{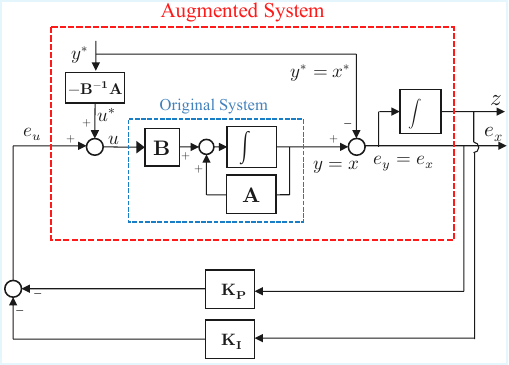}
  \caption{Schematic of the MIMO controller in state space form.}
  \label{fig:Augmented}
\end{figure}

\begin{figure}[!t]
  \centering
  \includegraphics[width=\linewidth]{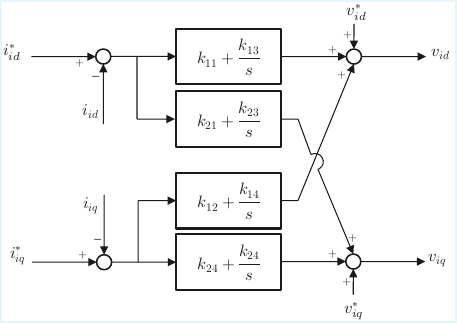}
  \caption{Schematic of the developed current controller for VSCs.}
  \label{fig:MIMO_PI}
\end{figure}

This section explains the developed current controller for VSCs, which is based on optimal control theory \cite{Wong1985, Puleston1993}. The state space model in \eqref{eq:diid/dt} and \eqref{eq:diiq/dt} can be represented as follows,
\begin{equation}
\label{eq:ss_model}
\begin{aligned}
\dot{x} &= \mathbf{A}\,x + \mathbf{B}\,u\\
y &= \mathbf{C}\,x
\end{aligned}
\end{equation}
\noindent\begingroup
\setlength{\arraycolsep}{3pt}
\renewcommand{\arraystretch}{0.95}
where $\mathbf A=\!\begin{bmatrix}
-\frac{R_f}{L_f} & \omega\\[-1pt]
-\omega & -\frac{R_f}{L_f}
\end{bmatrix},\,
\mathbf B=\!\begin{bmatrix}
\frac{1}{L_f} & 0\\[-1pt]
0 & \frac{1}{L_f}
\end{bmatrix},\,
\mathbf C=\!\begin{bmatrix}
1 & 0\\[-1pt]
0 & 1
\end{bmatrix}.$
\endgroup
In this representation, \(x = y = [\,i_{id},\, i_{iq}\,]^{\mathsf{T}}\) and
\(u = [\,v_{id},\, v_{iq}\,]^{\mathsf{T}}\).
In order to keep the system states at a desired setpoint
\(x^{*} = [\,i_{id}^{*},\, i_{iq}^{*}\,]^{\mathsf{T}}\), one needs a constant input
\(u^{*} = [\,v_{id}^{*},\, v_{iq}^{*}\,]^{\mathsf{T}}\) that satisfies \eqref{eq:ss_equilibrium}. 
\begin{equation}
\label{eq:ss_equilibrium}
0 = \mathbf{A}\, x^{*} + \mathbf{B}\, u^{*}
\end{equation}
Since $\mathbf{B}$ is invertible, $u^{*}$ is calculated as \eqref{eq:u_star}.
\begin{equation}
\label{eq:u_star}
u^{*} = -\,\mathbf{B}^{-1}\mathbf{A}\,x^{*}
\end{equation}
According to \cite{Kwakernaak1972}, the error terms in the input, states and the output of the system are defined as in \eqref{eq:e_x} and \eqref{eq:e_u}, respectively.
\begin{equation}
\label{eq:e_x}
e_x = e_y = x - x^{*} =
\begin{bmatrix}
i_{id} - i_{id}^{*} \\
i_{iq} - i_{iq}^{*}
\end{bmatrix}
\end{equation}
\begin{equation}
\label{eq:e_u}
e_u = u - u^{*} =
\begin{bmatrix}
v_{id} - v_{id}^{*} \\
v_{iq} - v_{iq}^{*}
\end{bmatrix}
\end{equation}
The integral terms in the PI controller are introduced as new states called $z(t)$ as in \eqref{eq:z_int}.
\begin{equation}
\label{eq:z_int}
z(t) = \int_{0}^{t} \big[x(\tau) - x^{*}\big]\,d\tau
\end{equation}
A new state space with the addition of the integral terms is formed. This augmented state space is as in \eqref{eq:aug_vars} and \eqref{eq:aug_ss}, respectively.
\begin{equation}
\label{eq:aug_vars}
\bar{x} = \bar{y} =
\begin{bmatrix}
e_x \\
z
\end{bmatrix}
\end{equation}
\begin{equation}
\label{eq:aug_ss}
\begin{aligned}
\dot{\bar{x}} &= \bar{\mathbf{A}}\,\bar{x} + \bar{\mathbf{B}}\,e_u,\\
\bar{y}       &= \bar{\mathbf{C}}\,\bar{x}.
\end{aligned}
\end{equation}
where
\begin{equation}
\label{eq:Abar_def}
\bar{\mathbf{A}} =
\begin{bmatrix}
\mathbf{A} & \mathbf{0}\\[2pt]
\mathbf{C} & \mathbf{0}
\end{bmatrix}_{4\times 4}
\end{equation}
\begin{equation}
\label{eq:Bbar_def}
\bar{\mathbf{B}} =
\begin{bmatrix}
\mathbf{B}\\[2pt]
\mathbf{0}
\end{bmatrix}_{4\times 2}
\end{equation}
\begin{equation}
\label{eq:Cbar_def}
\bar{\mathbf{C}} =
\begin{bmatrix}
\mathbf{C} & \mathbf{0}\\[2pt]
\mathbf{0} & \mathbf{I}
\end{bmatrix}_{4\times 4}
\end{equation}
Once the augmented system states in \eqref{eq:aug_vars} go to zero, the original system will also track the setpoint $x^*$. Given the structure of the augmented system it is straightforward show that controllability matrix associated with the augmented system \eqref{eq:ctrb_aug} is full rank, hence the augmented system is controllable.
\begin{equation}
\label{eq:ctrb_aug}
\mathcal{C} =
\big[\,\bar{\mathbf{B}}\;\; \bar{\mathbf{A}}\bar{\mathbf{B}}\;\; \bar{\mathbf{A}}^{2}\bar{\mathbf{B}}\;\; \bar{\mathbf{A}}^{3}\bar{\mathbf{B}}\,\big]
\end{equation}
The objective of the controller is to minimize the performance measure defined in \eqref{eq:lqr_cost}.
\begin{equation}
\label{eq:lqr_cost}
J = \tfrac{1}{2}\int_{0}^{\infty}
\big(\bar{x}^{\top}\bar{\mathbf{Q}}\,\bar{x} + e_u^{\top}\bar{\mathbf{R}}\,e_u\big)\,dt
\end{equation}
In \eqref{eq:lqr_cost}, $\bar{\mathbf{R}}$ represents the input penalties and it is chosen to be a positive definite matrix. $\bar{\mathbf{Q}}$ represents the state penalties and is selected to be a positive semidefinite matrix. According to \cite{Puleston1993}, the optimal input is achieved as in \eqref{eq:optimal_law}, 
\begin{equation}
\label{eq:optimal_law}
e_u = -\,\bar{\mathbf{R}}^{-1}\bar{\mathbf{B}}^{\top}\mathbf{P}\,\bar{x}
      = -\,\mathbf{K}\,\bar{x}
\end{equation}
where $\bar{\mathbf{P}}$ is obtained by solving the algebraic Riccati equation as shown in \eqref{eq:care} and $\mathbf{K}$ is a $2\times4$ matrix as in \eqref{eq:K_entries}.
\begin{equation}
\label{eq:care}
\mathbf{P}\,\bar{\mathbf{B}}\,\bar{\mathbf{R}}^{-1}\bar{\mathbf{B}}^{\top}\mathbf{P}
- \mathbf{P}\,\bar{\mathbf{A}} - \bar{\mathbf{A}}^{\top}\mathbf{P}
- \bar{\mathbf{C}}^{\top}\bar{\mathbf{Q}}\,\bar{\mathbf{C}} = \mathbf{0}
\end{equation}
\begin{equation}
\label{eq:K_entries}
\mathbf{K} =
\begin{bmatrix}
k_{11} & k_{12} & k_{13} & k_{14}\\
k_{21} & k_{22} & k_{23} & k_{24}
\end{bmatrix}
\end{equation}
By choosing proper penalties, one can find a desirable response by finding a tradeoff between the control input effort $e_u$ and the speed at which the new states $e_{x}(t)$ and $z(t)$ go to zero. The advantage of this method is that the penalties, specifically $\bar{\mathbf{Q}}$, can be selected to distinguish between the priority given to the integral gains or proportional gains of the PI controller. In addition, one can prioritize the response of one axis to achieve better transient and steady-state tracking specifications than the other axis.
\begin{equation}
\label{eq:pi_law}
u = \mathbf{K}_{\mathrm{P}}\,(x^{*}-x)
  + \mathbf{K}_{\mathrm{I}}\!\int_{0}^{t}\!\big[x^{*}-x(\tau)\big]\,d\tau
  + u^{*}
\end{equation}
where the proportional and integral controller gains are as in \eqref{eq:KP_block} and \eqref{eq:KI_block}, respectively. 
\begin{equation}
\label{eq:KP_block}
\mathbf{K}_{\mathrm{P}} =
\begin{bmatrix}
k_{11} & k_{12}\\
k_{21} & k_{22}
\end{bmatrix}
\end{equation}
\begin{equation}
\label{eq:KI_block}
\mathbf{K}_{\mathrm{I}} =
\begin{bmatrix}
k_{13} & k_{14}\\
k_{23} & k_{24}
\end{bmatrix}
\end{equation}
\cref{fig:Augmented} shows the schematic of the augmented system and the controller based on the state space form. \cref{fig:MIMO_PI} shows the current controller according to \eqref{eq:pi_law}. As shown in \cref{fig:MIMO_PI}, its structure closely resembles that of conventional PI controllers shown in \cref{fig:SISO_PI}. As shown in the next section, the MIMO controller significantly improves dynamic performance and grid synchronization stability of IBRs in various conditions. 
\section{Case Study Results}
The system, as depicted in \cref{fig:GFL}, is simulated with the parameters listed in Table I. The delay of $t_d=1.5/f_{sw}$ is considered in the control loops due to one switching period for the discrete transformation in the digital controller, and half a sample for the PWM operation \cite{Freijedo2015}. Several scenarios are studied to demonstrate the advantages of the developed current controller. The controller parameters remain unchanged during these scenarios. Additionally, small-signal stability analysis provides an analytical demonstration of the controller's merits. The following subsections discuss the details of the studied cases. 
\subsection{Step changes on the \texorpdfstring{$d$}{d}-axis set point}
\begin{figure}[!t]
  \centering
  \includegraphics[width=\linewidth]{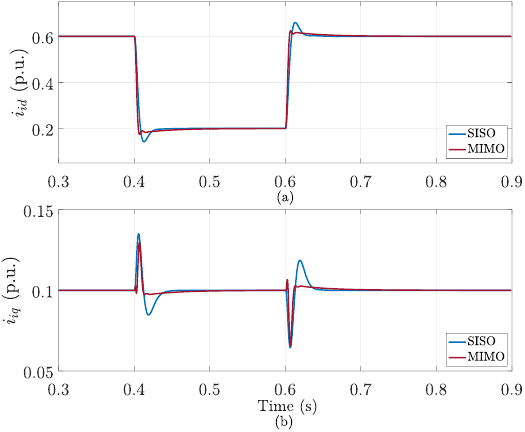}
  \caption{Comparison of controllers’ transient responses (a) $d$-axis (b) $q$-axis in facing the change in $d$-axis current reference in a relatively strong grid condition $\mathrm{SCR}=5$.}
  \label{fig:5}
\end{figure}

First, the set points of the $d$ and $q$ axes are changed individually to show the system's response to various step-up and step-down changes to the set points. 
\begin{table}[t]
  \caption{Test System Parameters}
  \label{tab:test-system-params}
  \centering
  \renewcommand{\arraystretch}{1.15}
  \begin{tabular}{@{}clc@{}}
    \toprule
    \textbf{Symbol} & \textbf{Description} & \textbf{Value} \\
    \doublemidrule
    $R_f$ & Resistance of the LC filter & $20\,\mathrm{m}\Omega$ \\
    $L_f$ & Inductance of the filter & $600\,\mu\mathrm{H}$ \\
    $C_f$ & Capacitance of the filter & $12\,\mu\mathrm{F}$ \\
    $f_{sw}$ & Switching frequency of the IBR & $5000\,\mathrm{Hz}$ \\
    $v_g$ & Nominal voltage of the grid & $500\,\mathrm{V}$ \\
    $S_{\mathrm{IBR}}$ & Nominal capacity of the IBR & $100\,\mathrm{kW}$ \\
    $k_p$ & Proportional gain of the PI current controller & $0.13$ \\
    $k_i$ & Integral gain of the PI current controller & $11.25$ \\
    $k_p^{PLL}$ & Proportional gain of the PLL & $48$ \\
    $k_i^{PLL}$ & Integral gain of the PLL & $144$ \\
    $t_d$ & Control delay & $\dfrac{3}{2\,f_{sw}}$ \\ \\
    \doublemidrule
    \multicolumn{3}{c}{\textit{Parameters of developed MIMO current controller}}\\
    \doublemidrule
    \multicolumn{3}{c}{
      $\bar{\mathbf{Q}} =
      \begin{bmatrix}
        0.0769 & 0 & 0 & 0\\
        0 & 0.0769 & 0 & 0\\
        0 & 0 & 70 & 0\\
        0 & 0 & 0 & 70
      \end{bmatrix} \quad
      \bar{\mathbf{R}} =
      \begin{bmatrix}
        1 & 0\\
        0 & 1
      \end{bmatrix}$} \\[0.4em]
    \multicolumn{3}{c}{
      $\mathbf{K}_{\mathrm{P}} =
      \begin{bmatrix}
        0.2690 & 0\\
        0 & 0.2690
      \end{bmatrix} \quad
      \mathbf{K}_{\mathrm{I}} =
      \begin{bmatrix}
        7.0076 & -4.5710\\
        4.5710 & 7.0076
      \end{bmatrix}$} \\
    \bottomrule
  \end{tabular}
\end{table}
Initially, the grid is relatively strong with $\mathrm{SCR}=5$ at the point of common coupling of the IBR to the grid, with $X_g/R_g=\tan(80^\circ)$. First, the current setpoint of the $d$-axis is changed from $0.6$ p.u. to $0.2$ p.u. and back to $0.6$ p.u. at $t=0.4$ $s$ and $t=0.6$ $s$, respectively, while the $q$-axis setpoint is fixed at $0.1$ p.u. As shown in \cref{fig:5}, the developed MIMO controller is noticeably superior, with less oscillation and faster tracking response. While it is $2$ $ms$ faster, it has $9\%$ less overshoot, and its $5\%$ settling time is $3.3$ $ms$ faster. In addition, the developed controller provides a better decoupling of the axes; while the $d$-axis is changed, the $q$-axis is affected much less than the conventional SISO PI. 

\begin{figure}[!t]
  \centering
  \includegraphics[width=\linewidth]{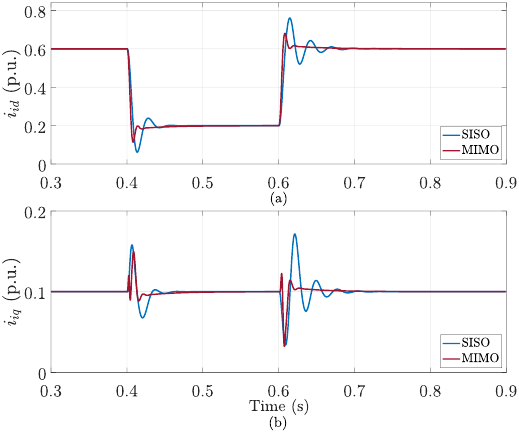}
  \caption{Comparison of controllers’ transient responses (a) $d$-axis (b) $q$-axis in facing the change in $d$-axis current reference in an extremely weak grid condition ($\mathrm{SCR}=1.95$).}
  \label{fig:6}
\end{figure}

The same scenario of changing the setpoint is investigated, but the $\mathrm{SCR}$ is $1.95$, representing an extremely weak grid. As shown in \cref{fig:6}, the MIMO controller can maintain a desirable response, while the SISO controller oscillates noticeably. PLL dynamics combined with the weak grid’s impedance create a strong coupling with the current controller loop \cite{Wen2016}. As a result, the ideal proportional decoupling terms in conventional controllers fail to cancel out the coupling entirely. On the other hand, the MIMO controller successfully damps out these oscillations. In summary, the MIMO controller provides superior dynamic performance in both strong and weak grid conditions and better decoupling of the controller's two axes.
\subsection{Step changes on the \texorpdfstring{$q$}{q}-axis set point}
\begin{figure}[!t]
  \centering
  \includegraphics[width=\linewidth]{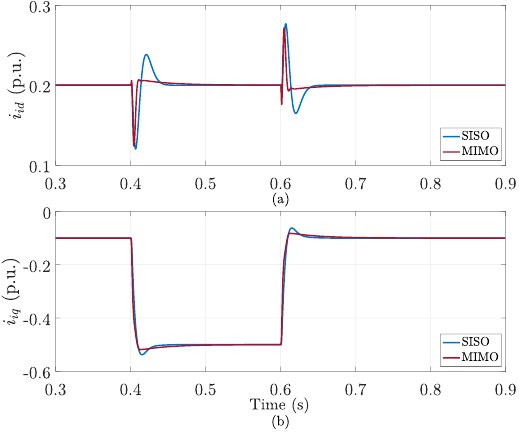}
  \caption{Comparison of controllers’ transient responses (a) $d$-axis (b) $q$-axis in facing the change in $q$-axis current reference in a relatively strong grid condition $\mathrm{SCR}=5$.}
  \label{fig:7}
\end{figure}

\begin{figure}[!t]
  \centering
  \includegraphics[width=\linewidth]{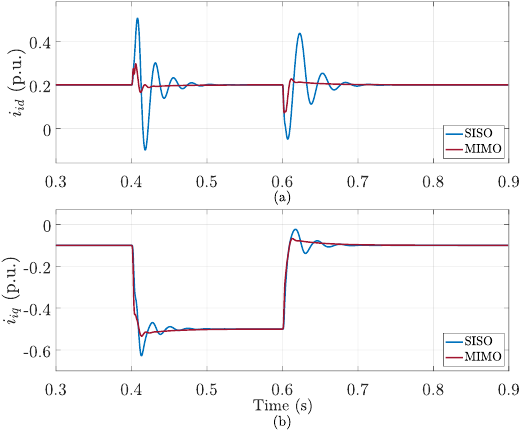}
  \caption{Comparison of controllers’ transient responses (a) $d$-axis (b) $q$-axis in facing the change in $q$-axis current reference in an extremely weak grid condition ($\mathrm{SCR}=1.95$).}
  \label{fig:8}
\end{figure}

The $q$-axis is directly responsible for reactive power control, which is a crucial factor in maintaining voltage stability in power grids. The injected power to the grid will deviate from the desired setpoints if this response is sluggish, and if oscillatory, the system will be prone to undamped oscillations that may lead to loss of synchronism of the IBR. Additionally, better dynamic response of $i_{iq}$ helps to counteract fluctuations in the voltage, allowing the faster PLL to track the changes in the grid. \cref{fig:7} and \cref{fig:8} show the performance of each controller when a step change is applied on the $q$-axis in a strong and weak grid, respectively. As shown in these figures, the MIMO controller has a noticeably superior dynamic performance in both cases. The performance improvement of the MIMO controller is more evident in the weak grid scenario. 
\subsection{The effect of impedance angles}
\begin{figure}[!t]
  \centering
  \includegraphics[width=\linewidth]{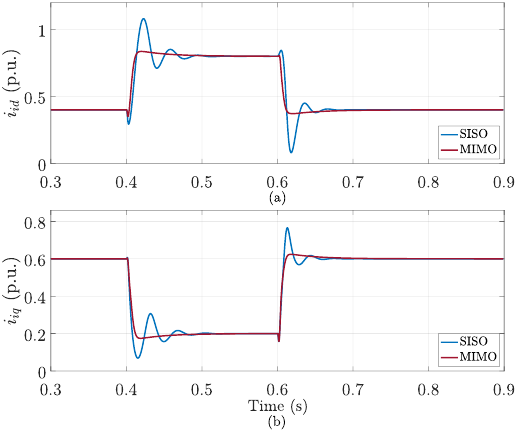}
  \caption{Comparison of controllers’ transient responses (a) $d$-axis (b) $q$-axis in facing the changes on both $d$ and $q$ axes in an extremely weak grid ($\mathrm{SCR}=1$) with $X_g/R_g=1$.}
  \label{fig:9}
\end{figure}

Equations \eqref{eq:P_inv} and \eqref{eq:Q_inv} represent the static power transfer capability of a grid tied VSC, assuming the inverter has an output voltage $v_o$ with the angle $\delta$ \cite{Zhou2012IET}.
\begin{equation}
\label{eq:P_inv}
P_{\mathrm{inv}}
= \frac{v_o v_g}{\lvert Z_g\rvert}
  \sin\!\Big(\delta - 90^{\circ} + \tan^{-1}\!\frac{X_g}{R_g}\Big)
  + v_o^{2}\,\frac{R_g}{\lvert Z_g\rvert^{2}}
\end{equation}
\begin{equation}
\label{eq:Q_inv}
Q_{\mathrm{inv}}
= -\,\frac{v_o v_g}{\lvert Z_g\rvert}
   \cos\!\Big(\delta - 90^{\circ} + \tan^{-1}\!\frac{X_g}{R_g}\Big)
   + v_o^{2}\,\frac{X_g}{\lvert Z_g\rvert^{2}}
\end{equation}
Therefore, the static limit of the active power of the grid tied VSC in the p.u. system is shown in \eqref{eq:Ppu_max}. In this condition the injected reactive power is as \eqref{eq:Qpu}.
\begin{equation}
\label{eq:Ppu_max}
P^{\max}_{p.u}
= \mathrm{SCR}\,\Big(\frac{v_g}{v_o} + \frac{R_g}{\lvert Z_g\rvert}\Big)
\end{equation}
\begin{equation}
\label{eq:Qpu}
Q_{p.u}
= \mathrm{SCR}\,\frac{X_g}{\lvert Z_g\rvert}
\end{equation}
where the SCR is defined as in \eqref{eq:SCR_def}.
\begin{equation}
\label{eq:SCR_def}
\mathrm{SCR}
= \frac{v_o^{2}}{\lvert Z_g\rvert\,S_{\mathrm{IBR}}}
\end{equation}
In the previous section, $X_g/R_g=\tan(80^\circ)$ was considered. However, some reports have investigated weak grid incidents  with a low $X_g/R_g$ ratio as well \cite{Harasis2021}. Hence, this section investigates the controller's performance in low-impedance angles, where $X_g/R_g=1$. \cref{fig:9} depicts the response of the controller with this impedance ratio. Assuming the magnitude of $v_o$ is close to $1$ p.u., according to \eqref{eq:Ppu_max}, the static power transfer limit is $\mathrm{SCR}\times1.707$ p.u. However, due to dynamic limits, at $\mathrm{SCR}=1$, the SISO controller can inject at most $0.94$ p.u. of active power before it loses its grid synchronization and becomes unstable in the studied system. The MIMO controller can inject $1.66$ p.u. of active power, which means the active power transfer capability has improved by $36\%$. This is because the MIMO controller provides better dynamic performance, as shown in Fig. 9. Also, at $\mathrm{SCR}=1$, the SISO controller can inject at most $0.49$ p.u. of reactive power, but the MIMO controller can go as high as $0.66$ p.u. of reactive power. which is $24\%$ higher than that of the conventional SISO controller.
\subsection{Grid synchronization stability study using time-domain simulation}
\begin{figure}[!t]
  \centering
  \includegraphics[width=\linewidth]{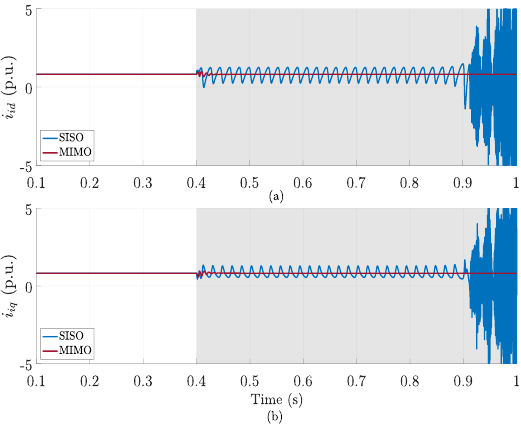}

  \vspace{2pt} 

  \includegraphics[width=\linewidth]{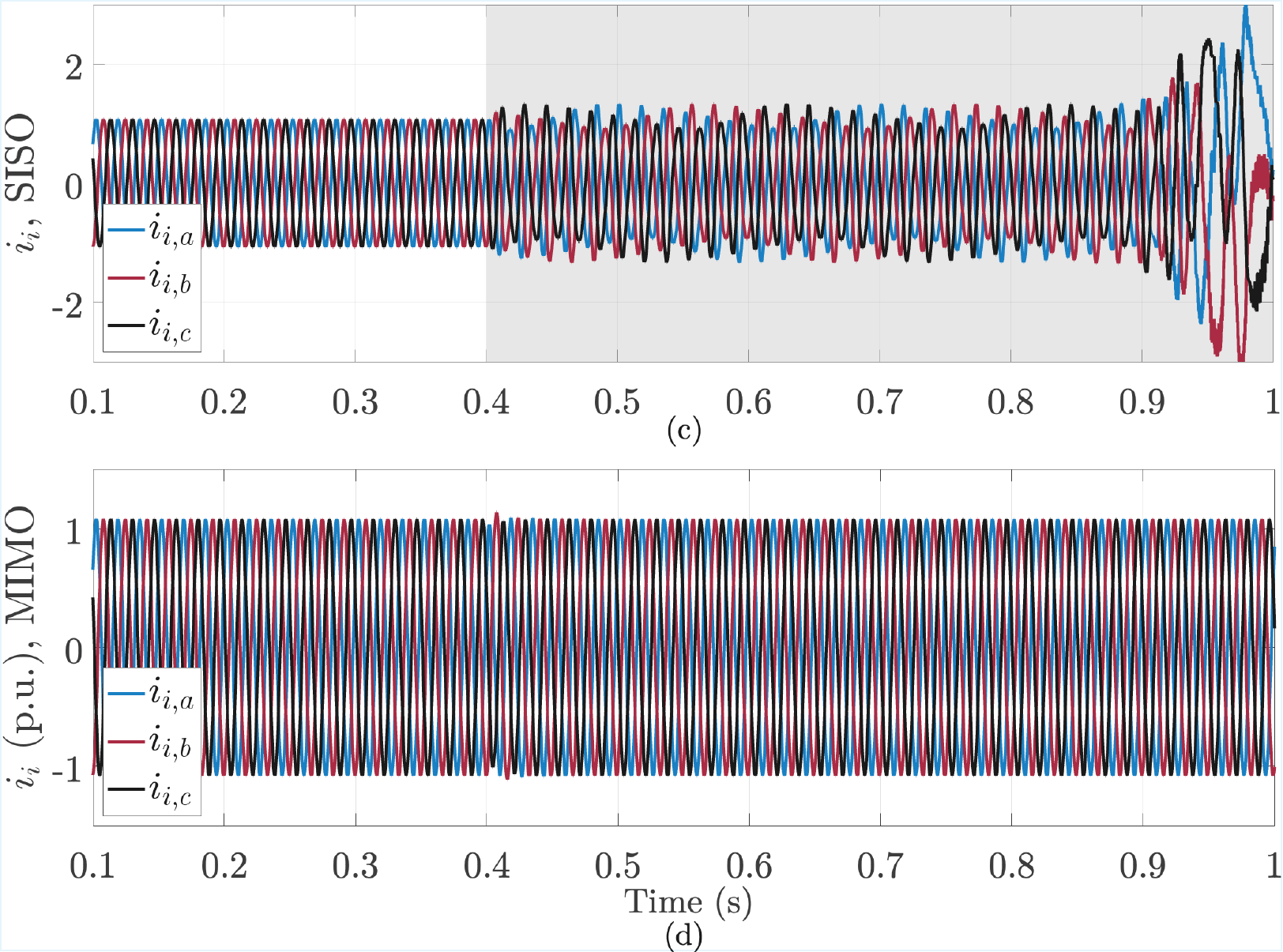}

  \caption{Comparison of the controllers’ responses to a sudden change to the grid strength changing from $\mathrm{SCR}=4$ to $\mathrm{SCR}=2$ at $t=0.4$ $s$; (a) and (b) represent the currents in SRF, (c) and (d) represent the current of the SISO and MIMO in $abc$ frame, respectively.}
  \label{fig:10}
\end{figure}

This section demonstrates the enhancement of grid synchronization stability by the MIMO controller in response to sudden changes in grid strength. This scenario is motivated by a real-life incident in a solar power plant in Arizona \cite{Morjaria2017}. In that incident, due to outages of a few lines, the strength of the grid dropped, causing oscillatory power injections by the solar power plant. In this subsection, it is assumed that the VSC is interfaced to the grid through two identical transmission lines. Initially, the grid strength is $\mathrm{SCR}=4$, with an impedance angle of $80$ degrees. The VSC injects approximately equal active and reactive powers of $0.66$ p.u. This is well below the static limit of $P_{p.u}^{max}=4.69$ p.u. and its corresponding reactive power of $Q_{p.u}=1.67$ p.u., calculated based on \eqref{eq:Ppu_max} and \eqref{eq:Qpu}. At $t=0.4$ $s$, one of the lines is taken out, dropping the $\mathrm{SCR}$ to $2$. Following this event, as shown in \cref{fig:10}-(c), the SISO controller fails to synchronize to the system. On the other hand, as shown in \cref{fig:10}-(d), the MIMO controller maintains synchronism, which shows its superior ability to maintain the grid synchronization stability. 
\subsection{Grid synchronization stability study using analytical analysis}
\begin{figure}[!t]
  \centering
  \includegraphics[width=\linewidth]{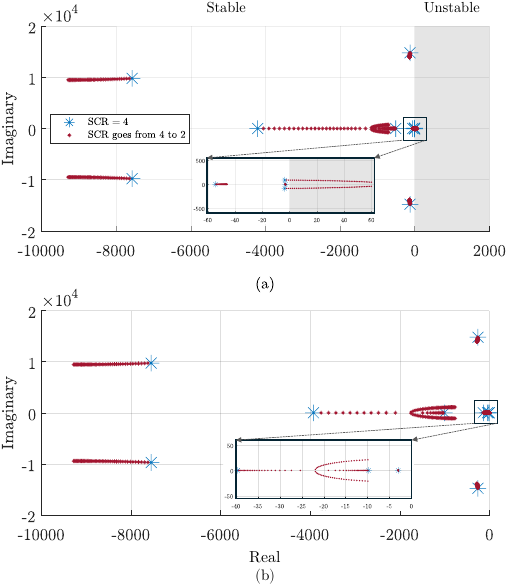}
  \caption{Comparison of eigenvalues of the system with (a) SISO-PI (b) MIMO-PI controllers at different grid strength values.}
  \label{fig:11}
\end{figure}

In this section, eigenvalue analysis is performed to study the grid synchronization stability enhancement by the MIMO PI controller. The state space of the grid-tied system with each configuration is derived according to \cite{Li2019Energies}. \cref{fig:11} shows the system's eigenvalues for systems with different grid strength, where \cref{fig:11}-(a) shows the results for the conventional SISO controller and \cref{fig:11}-(b) shows the results for the MIMO controller. The system eigenvalues are plotted by sweeping through strength values from $\mathrm{SCR}=4$ to $\mathrm{SCR}=2$, observing their corresponding changes. It is evident that for the SISO controller, two of the eigenvalues soon move towards the right half-plane, causing the system to become unstable. In contrast, the MIMO controller remains stable even under weaker conditions.

\begin{figure}[!t]
  \centering
  \includegraphics[width=\linewidth]{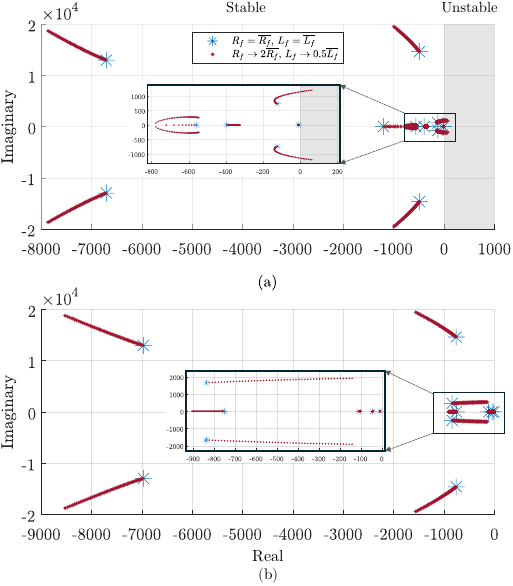}
  \caption{Comparison of the eigenvalues in (a) SISO-PI and (b) MIMO-PI controllers by changing the filter parameters of the VSC.}
  \label{fig:12}
\end{figure}

The grid synchronization stability of the VSC under parameter variations is also investigated. Fig.12 shows the eigenvalues for $50$ different values of the resistor and inductor of the filter. These values are selected from a range that spans from the nominal value used in the controller design to twice the nominal value for the resistor and half the nominal value for the inductor. The grid strength is set at $\mathrm{SCR}=2$. As shown in \cref{fig:12}, the eigenvalues in the MIMO-PI controller move toward the right half-plane at a slower rate compared to those in the SISO controller.

\section{Conclusion}
This paper developed a current controller strategy based on optimal control theory that has a structure that closely resembles that of commonly used controllers in VSC. It can be readily implemented without adding complexity to the controller structure. By formulating the control design as an optimization problem, it was shown that the LQR naturally selects only the integral action, in contrast to conventional SISO PI controllers, which rely on proportional terms to achieve decoupling. The developed controller provides superior dynamic performance in both strong and weak power grid conditions. Time domain simulation results and analytical analysis demonstrated enhanced grid synchronization stability and power transfer capability of the developed controller.

\bibliographystyle{IEEEtran}
\bibliography{bare_jrnl.bib}
\ifCLASSOPTIONcaptionsoff
  \newpage
\fi
\vspace{4.5cm} 

\begin{IEEEbiography}[
  {\includegraphics[width=1in,height=1.25in,clip,keepaspectratio]{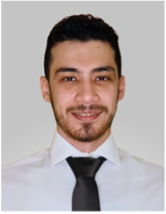}}
]{Hassan Yazdani}
(S’25) is currently pursuing a Ph.D. in Electrical Engineering at Washington State University, Pullman, WA, USA. His research interests include the dynamics, control, and stability analysis of inverter-based power systems, as well as optimization and machine learning applications in power grids.
\end{IEEEbiography}

\vspace{-9.5cm} 

\begin{IEEEbiography}[
  {\includegraphics[width=1in,height=1.25in,clip,keepaspectratio]{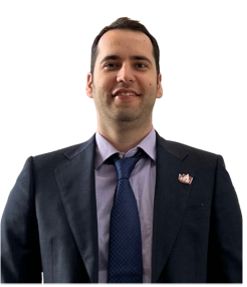}}
]{Ali Maleki}
received his B.Sc. degree in Electrical Engineering from K. N. Toosi University of Technology, Tehran, Iran, in 2016 and his M.Sc. degree in Electrical Engineering from Tehran Polytechnic, Tehran, Iran, in 2019. He is currently pursuing a Ph.D. in the School of Electrical Engineering and Computer Science at Washington State University, Pullman, WA, USA. His research interests include renewable energy integration, electricity markets, and control theory.
\end{IEEEbiography}

\vspace{-9.5cm} 

\begin{IEEEbiography}[
  {\includegraphics[width=1in,height=1.25in,clip,keepaspectratio]{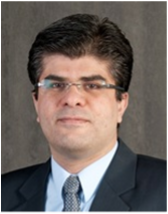}}
]{Saeed Lotfifard}
(S’08–M’11–SM’17) received his Ph.D. degree in electrical engineering from Texas A\&M University, College Station, TX, in 2011. Currently, he is an associate professor at Washington State University, Pullman. His research interests include stability, protection, and control of inverter-based power grids. He serves as an associate editor for the \textsc{IEEE Transactions on Power Delivery}.
\end{IEEEbiography}

\vspace{-9.2cm} 

\begin{IEEEbiography}[
  {\includegraphics[width=1in,height=1.25in,clip,keepaspectratio]{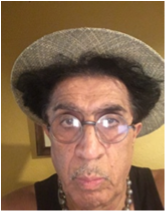}}
]{Ali Saberi}
lives and works in Pullman, WA, USA.
\end{IEEEbiography}
\end{document}